# Surface plasmon resonance of two-segmented Au–Cu nanowires in polycarbonate template

F. Babaei[*], A. Azarian

Department of Physics, University of Qom, Qom, Iran
[*]) E-mail: fbabaei@qom.ac.ir

## Abstract

Two segmented gold- copper nanowires were grown inside the pores of polycarbonate track (PCT) etched membranes from two separate solutions by electrodeposition method. Optical absorption spectra of two segmented Au-Cu nanowires in PCT template were shown a surface plasmon resonance peak at about 900 nm for incident angle of $\Theta = 65$ degrees but for $\Theta = 0$ there are not any peaks in spectra.

## 1. Introduction

In recent years, two segmented nanowires has generated wide interest because of their shape - dependent optoelectronic properties and their potential applications in magnetic devices, photonic and catalysis and nanobarcodes [1, 2]. One of the great interests in current research is surface plasmon resonance (SPR) of noble metals nanowires. Because, SPR depends on dimension of the nanowire and provide tunable optical characteristics for various application such as nanobarcodes. Au is particularly attractive material in multi segments structures because of stability, sensibility in biochemistry and large difference in reflectivity between gold and other metallic segments which it provides excellent contrast under visible illumination for readout of striping pattern in nanobarcodes. In spite of various works which have been done on SPR of nanostructures, to the best of our knowledge, this is the first work that study SPR properties of two segmented nanowires in PCT template.

In our earlier work [3] we reported the excitation of SPR of the Cu nanowires embedded in polycarbonate membrane. The aim of this work is to report on SPR in two segmented Au-Cu nanowires .Then we measured optical absorption spectra of these nanostructures in PCT template and compared with optical modeling on the base of transfer matrix method [4-6]. The fabrication method and optical modeling are outlined in Section 2 & 3, respectively. Results are presented and discussed in Section 4.

## 2. Experimental

In this work we prepared two segmented gold- copper nanowires inside the pores of 50 nm nominal diameter polycarbonate track etched membranes. Au and Cu materials were electrodeposited from two separate solutions. One could find details of experimental procedure in our previous work [3, 7]. In summery, Au were electrodeposited from solutions containing 100 mM $KAu(CN)_2$ and 0.25 M $Na_2CO_3$ at pH 13. Experiments were performed in a conventional three electrode cell and at a constant potential of -0.9 V vs. saturated calomel electrode (SCE). Cu was grown inside the pores by changing the previous solution by solutions 0.2M $CuSO_4.5\ H_2O$ and 0.4M $H_3BO_3$. All electrochemical experiments were performed with a potentiostat/galvanostat (Princeton Applied Research, model 263A) at room temperature. A Pt foil (2 $cm^2$) and a SCE served as the counter electrode and reference electrode, respectively. Prior to electrodeposition of the wires, back side of polycarbonate membranes was first coated with a 50 nm layer of gold by sputtering technique. Then, the templates were extensively rinsed and immersed in the deposition solution for at least 20 min to ensure that all pores were uniformly wet.

In this work, we first measured the choronoamperometry (I-t) curve and calibrate it for grow copper and gold to desired length for anyone and then the growth time of each segments was choice in such away that length of each segments about 2 μm.

### 3. Optical modeling

The Kretschmann configuration is a common experimental arrangement for the excitation and detection of surface plasmon waves [8, 9]. In order to excitation surface plasmon, we used a configuration similar to that Kretschmann configuration is as follows: the region $0 \leq z \leq d_1$ is occupied by a metal of relative permittivity $\varepsilon_{met}$, the region $d_1 \leq z \leq d_2 + d_3$ is described by Cu - coated Au nanowires (the angle of rise of nanowire relative to substrate is $\chi$) embedded in polycarbonate membrane, whereas the regions $z \leq 0$ and $z \geq (d_1 + d_2 + d_3)$ are homogeneous isotropic dielectric material of relative permittivity $\varepsilon_l = n_l^2$, as it is shown in Fig.1.

A plane wave in the half space $z \leq 0$ propagate at an angle $\theta$ to the z- axis and at an angle $\psi$ to the x- axis in the xy - plane. The phasors of incident, reflected and transmitted electric fields are given as [4]:

$$\begin{cases} \underline{E}_{inc}(\underline{r}) = [a_s \underline{S} + a_p \underline{P}_+] e^{i\underline{K}_0 n_l \cdot \underline{r}}, & z \leq 0 \\ \underline{E}_{ref}(\underline{r}) = [r_s \underline{S} + r_p \underline{P}_-] e^{-i\underline{K}_0 n_l \cdot \underline{r}}, & z \leq 0 \\ \underline{E}_{tr}(\underline{r}) = [t_s \underline{S} + t_p \underline{P}_+] e^{i\underline{K}_0 n_l \cdot \underline{r}} e^{i\underline{K}_0 n_l \cdot (\underline{r} - l_\Sigma \underline{u}_z)}, & z \geq (d_1 + d_2 + d_3) \end{cases} \quad (1)$$

The magnetic field's phasor in any region is given as $\underline{H}(\underline{r}) = (i\omega\mu_0)^{-1} \underline{\nabla} \times \underline{E}(\underline{r})$ where $(a_s, a_p)$, $(r_s, r_p)$ and $(t_s, t_p)$ are the amplitudes of incident plane wave, and reflected and transmitted waves with S- or P- polarizations. We also have:

$$\begin{cases} \underline{r} = x\underline{u}_x + y\underline{u}_y + z\underline{u}_z \\ \underline{K}_0 = K_0(\sin\theta\cos\psi\ \underline{u}_x + \sin\theta\sin\psi\ \underline{u}_y + \cos\theta\ \underline{u}_z) \end{cases} \quad (2)$$

where $K_0 = \omega\sqrt{\mu_0\varepsilon_0} = 2\pi/\lambda_0$ is the free space wave number, $\lambda_0$ is the free space wavelength, $\varepsilon_0 = 8.854 \times 10^{-12}\ Fm^{-1}$ and $\mu_0 = 4\pi \times 10^{-7}\ Hm^{-1}$ are the permittivity and permeability of free space (vacuum), respectively. The unit vectors for linear polarization normal and parallel to the incident plane, $\underline{S}$ and $\underline{P}$, respectively are defined as:

$$\begin{cases} \underline{S} = -\sin\psi\ \underline{u}_x + \cos\psi\ \underline{u}_y \\ \underline{P}_{\pm} = \mp(\cos\theta\cos\psi\ \underline{u}_x + \cos\theta\sin\psi\ \underline{u}_y) + \sin\theta\ \underline{u}_z \end{cases} \quad (3)$$

and $\underline{u}_{x,y,z}$ are the unit vectors in Cartesian coordinates system.

The reflectance and transmittance amplitudes can be obtained, using the continuity of the tangential components of electrical and magnetic fields at interfaces and solving the algebraic matrix equation [5, 6]:

$$\begin{bmatrix} t_s \\ t_p \\ 0 \\ 0 \end{bmatrix} = [\underline{K}]^{-1} \cdot [\underline{M}_{Au-PCT}] \cdot [\underline{M}_{Cu-PCT}] \cdot [\underline{M}_{met}] \cdot [\underline{K}] \cdot \begin{bmatrix} a_s \\ a_p \\ r_s \\ r_p \end{bmatrix} \quad (4)$$

The different terms and parameters of this equation are given in detail by Lakhtakia and Messeir [6]. The reflection and transmission can be calculated as:

$$R_{i,j} = \left|\frac{r_i}{a_j}\right|^2,\quad T_{i,j} = \left|\frac{t_i}{a_j}\right|^2\ ;\quad i,j = s, p \quad (5)$$

## 4. Results and discussion

Scanning electron micrograph of nanowires after removal from the PCT templates by $CH_2Cl_2$ are shown in Fig.2. It is worth to note that image related to back scattering electron and hence the brighter part is related to the gold segment. It is clear that both copper and gold segments have length about 2 μm and diameter of wires is about 60 nm near nominal diameter of the PCT holes.

It is well known that the wavelength of maximum optical absorption depends on size, shape, optical properties of carrier medium, orientation and spacing between particles [10]. In two segmented structures, the distance between two inhomogeneous segments reach to zero. Therefore optical properties of these structures are different from optical properties of two separated systems. The optical absorption spectra of two segmented Au-Cu nanowires in PCT template at incident angles of light Ө = 0, 65° were shown in Fig. 3. It illustrates one peak locates at about 900 nm for Ө = 65 degrees but for Ө = 0 there are no peaks in spectrum.

For the purpose of simulation and comparison with experimental results, the structure is considered as a composite of Cu- Au coated inside the pores of PCT and a homogeneous gold thin film. In optical modeling, the relative permittivity scalars $\varepsilon_{a,b,c}$ of Cu coated Au nanowires in polycarbonate template were calculated using the Bruggeman homogenization formalism [11, 12]. In this formalism, the structure is considered as a two component composite (material deposition and polycarbonate). These quantities are dependent on different parameters, namely, columnar form factor, fraction of material deposition, the wavelength of free space and the refractive index $n(\lambda) + ik(\lambda)$. In addition, each column in the structure is considered as a string of identical long ellipsoids [6]. The ellipsoids are considered to be electrically small (i.e. small in a sense that their electrical interaction can be ignored) [13]. In all calculations columnar form factors $(\frac{c}{a})_{Cu\&Au} = (\frac{c}{a})_{PCT} = 20$, $(\frac{b}{a})_{Cu\&Au} = (\frac{b}{a})_{PCT} = 1.06$ (c, a, b are semi major axis ,

small half-axes of ellipsoids, respectively) [14] and experimental structural parameters $\chi = 90°$, $f_{Cu} = 0.008$ and $f_{Au} = 0.008$ in PCT were fixed. Setting the shape factors $(\frac{c}{a})_{Cu, Au \& PCT} \gg 1$ and $(\frac{b}{a})_{Cu, Au \& PCT} \gg 1$ will make each ellipsoid resemble a needle with a slight bulge in its middle part [6]. In our work, the absorbance for natural light $A = \frac{A_s + A_p}{2}$ as a function of $\theta$ is calculated where $A_i = 1 - \sum_{j=s,p} R_{ji} + T_{ji}$, $i = s, p$ (*linear ploarization*) [3] when $\psi = 0°$. The existence of sharp peak in optical absorbance of structure as angle of incident of light with linear polarization is due to excitation surface plasmon wave [15]. We have used the bulk experimental refractive indexes copper, gold and polycarbonate for homogenization [16]. Fig.4a shows that as the polar angle of incident light increases the SPR peak shifts to shorter wavelengths and also at first its intensity increases (until $\theta = 75°$) then decreases. We followed calculations for $\theta > 80°$ and found that the location of SPR does not changed (680 nm) but its intensity gradually can be neglected. Absorbance as function of wavelength is depicted in Fig 4b, when a surface plasmon wave has been excited ($\theta = 65°$) at different thicknesses $d_1$, $d_2$ and $d_3$. It can also be observed that as thickness of gold thin film ($d_1$) is zero surface plasmon wave can't be excited but for a 50 nm of gold thin film with $d_{Cu} = 4\mu m$, $d_3 = 0\mu m$ and $d_{Cu} = 0\mu m$, $d_{Au} = 4\mu m$ a SPR peak is respectively located at wavelength 864, 875nm. In order to simultaneously observe the influence of the incident polar angle and wavelength on surface plasmon wave propagation, plot3D and density plot of absorbance are depicted in Fig.5. Therefore, it can be seen that the results are the same. By comparing Figs. 3, 4 and 5, it is seen that there are qualitative agreements between experimental data and results of our used optical modeling.

## 4. Conclusions

Two segmented Au-Cu nanowires were grown within polycarbonate membrane having nominal pore size of 50 nm by electrodeposition method.. The results of experiment and optical modeling showed that a surface plasmon peak is appeared in $\theta_{inc} \approx 65^0$ at wavelength about 900 nm in our work. It shifts to shorter wavelengths as the incident of polar angle increases. Although, at higher polar angles its intensity decreases and the location of SPP peak is fixed at wavelength 680nm.This study may be applied to the plasmonic characterization of multi- segment of nanowires.

**Acknowledgements**

We wish to acknowledge support from the University of Qom.

**Figure captions**

Fig. 1. A schematic of the structure of thin film for optical modeling.

Fig.2. A typical SEM image of two segmented gold - copper nanowires on Si wafer after remove PCT template. The brighter part related to gold segment.

Fig.3. Optical absorption spectra of the nanowires embedded in PCT template at incident angles of light $\Theta = 0, 65°$.

Fig.4. a) Calculated absorbance as function of wavelength, when $\psi = 0^0$, at different $\theta$ for natural light plane wave. The structure is described by following parameters: $d_1 = 50nm$, $d_2 = 2\mu m$, $d_3 = 2\mu m$, $\varepsilon_l = 2.9$ (silicon oxynitride $SiO_{.4}N_{.6}$), $f_{Cu} = 0.008$, $f_{Au} = 0.008$ and $n_{PCT} = 1.584$, b) Calculated absorbance as function of wavelength, when a surface plasmon wave has been excited ($\theta = 65°$) for different thicknesses of $d_1$, $d_2$ and $d_3$.

Fig.5. a) Plot3D absorbance as function of wavelength and polar angle. b) Density plot of absorbance (see caption of Fig.4a for other parameters).

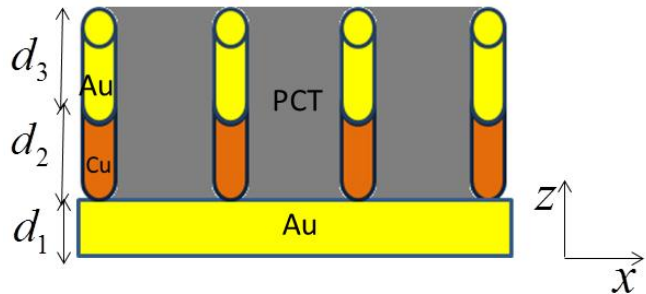

**Fig. 1; F. Babaei and A. Azarian**

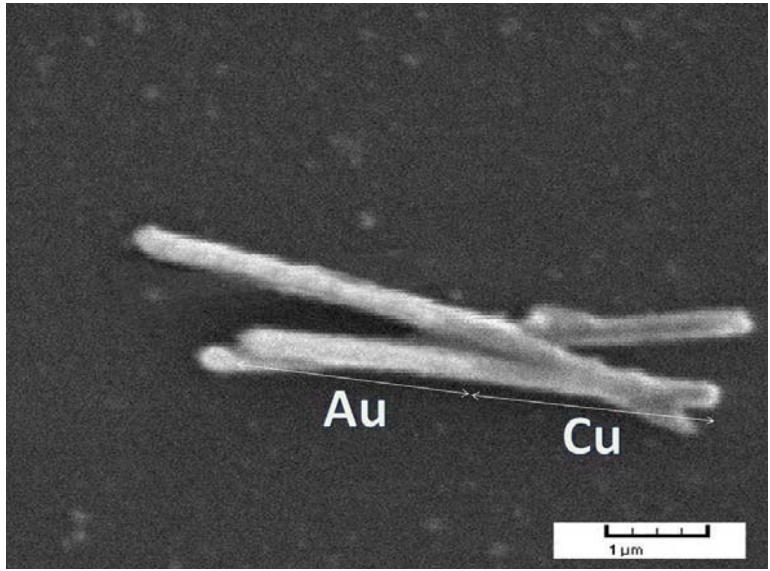

**Fig. 2; F. Babaei and A. Azarian**

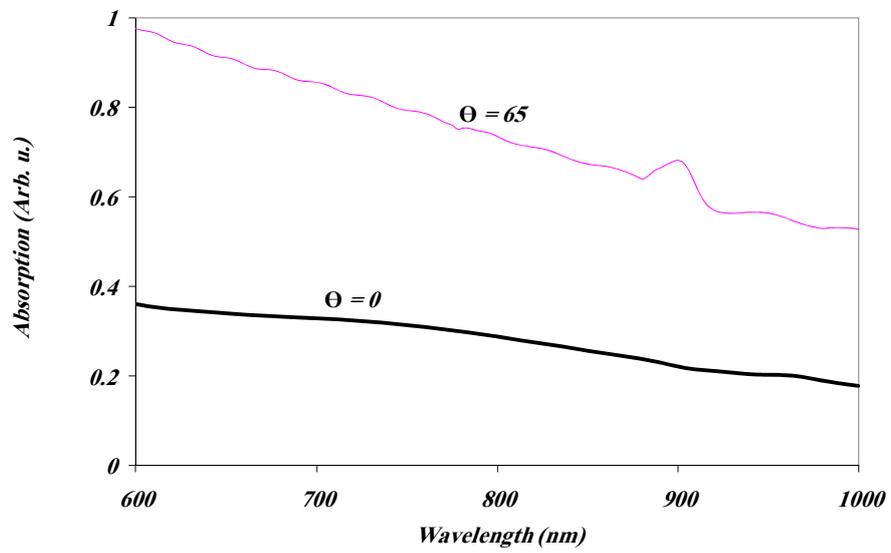

**Fig. 3; F. Babaei and A. Azarian**

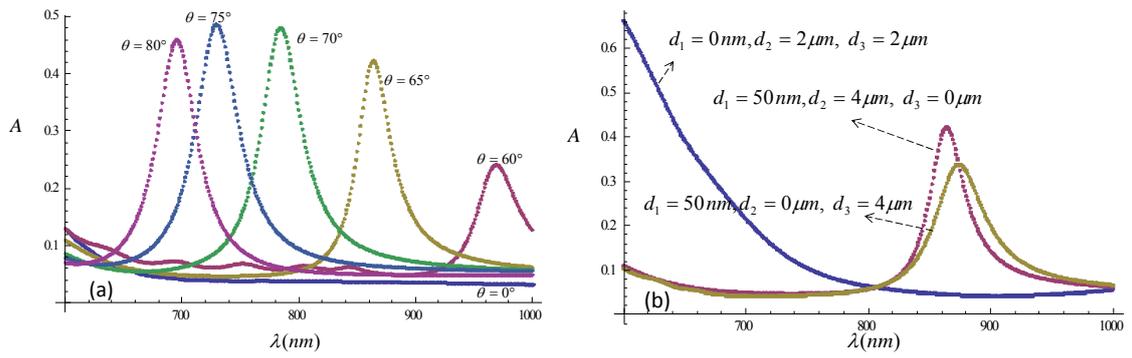

**Fig. 4; F. Babaei and A. Azarian**

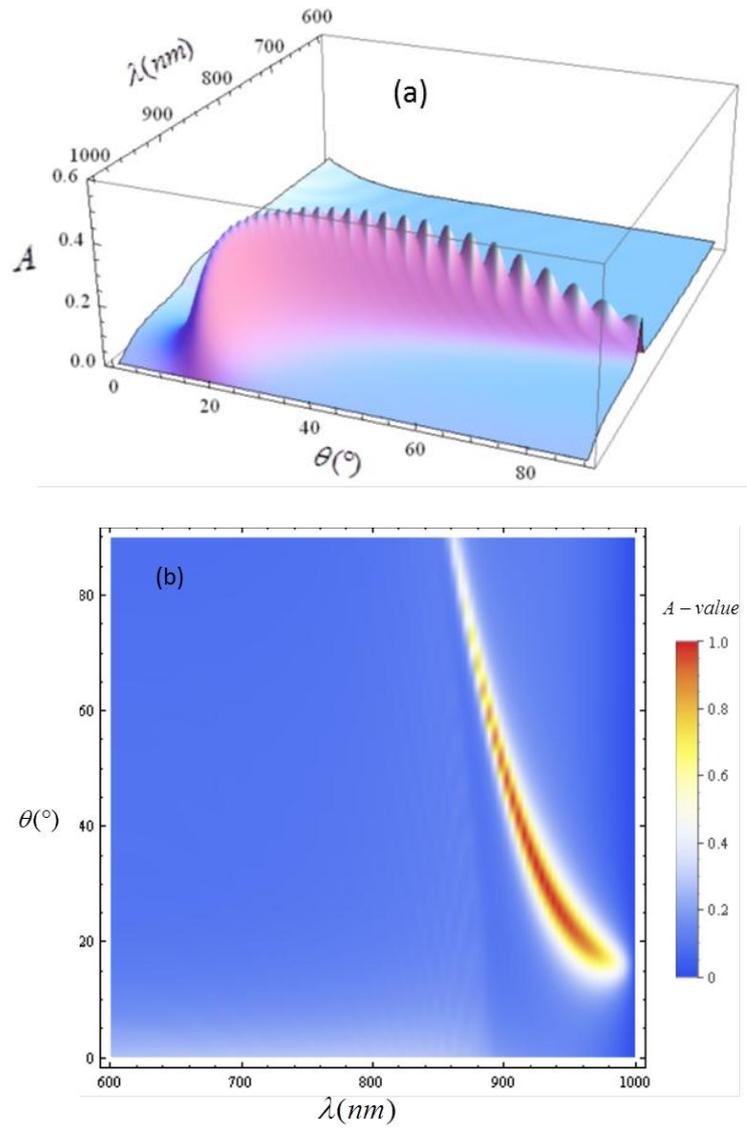

**Fig. 5; F. Babaei and A. Azarian**